
\input harvmac

\def\d#1{\{ #1 \}}

\def\p{^\prime}
\overfullrule=0pt

%
%
\def\RF#1#2{\if*#1\ref#1{#2.}\else#1\fi}
\def\NRF#1#2{\if*#1\nref#1{#2.}\fi}
\def\refdef#1#2#3{\def#1{*}\def#2{#3}}
%
%
\def \ts{\thinspace}

\def \NP{{\it Nucl.\ts Phys.\ts }}
\def \PL{{\it Phys.\ts Lett.\ts }}

\def \Tahoe{Proceedings of the NATO
 Conference on Differential Geometric Methods in Theoretical
 Physics, Lake Tahoe, USA 2-8 July 1989 (Plenum 1990) 169}
\def \Zam{Zamolodchikov}
\def \AZm{A.\ts B.\ts \Zam}
\def\dur{H.\ts W.\ts Braden, E.\ts Corrigan, P.\ts E.\ts Dorey and R.\ts
Sasaki}
\def\cds{E.\ts Corrigan, P.\ts E.\ts Dorey and R.\ts Sasaki}

\refdef\rAFZa\AFZa{A.\ts E.\ts Arinshtein, V.\ts A.\ts Fateev and
 \AZm,  \PL {\bf B87}
(1979) 389}

\refdef\rAKR\AKR{J. A. \ts Azc\'arraga, P. P. \ts Kulish and
F. \ts R\'odenas, Valencia preprint, FTUV 93-36}

\refdef\rBSa\BSa{H.\ts W.\ts Braden and R. Sasaki,  \PL {\bf B255} (1991) 343}

\refdef\rBSb\BSb{H.\ts W.\ts Braden and R. Sasaki,
\NP {\bf B379} (1992) 377}

\refdef\rBCDSa\BCDSa{\dur,
\PL {\bf B227} (1989) 411}

\refdef\rBCDSb\BCDSb{\dur, \Tahoe}

\refdef\rBCDSc\BCDSc{\dur, \NP {\bf B338} (1990) 689}

\refdef\rBCDSe\BCDSe{\dur,  \NP {\bf B356} (1991) 469}

\refdef\rCDS\CDS{\cds, \NP {\bf B408} (1993) 579}

\refdef\rCG \CG {E. \ts Cremmer  and J.-L. \ts Gervais,
 {\it Commun. Math. Phys. \bf 144} (1992) 279}

\refdef\rChe \Che {I. V.\ts Cherednik, {\it Teor. Matem. Fiz.} {\bf 61} (1984)
55}

\refdef\rCMa\CMa{P.\ts Christe and G.\ts Mussardo, \NP {\bf B330} (1990) 465}

\refdef\rCMb\CMb{P.\ts Christe and G.\ts Mussardo,
 {\it Int.~J.~Mod.~Phys.}~{\bf A5} (1990) 4581}

\refdef\rDeVR\DeVR{H.\ts J.\ts de\ts Vega and A.\ts Ruiz, {\it J. Phys.}
{\bf A26} (1978) L519; Preprint LPTHE 93-29}

 \refdef\rDGZc\DGZc{G. W. Delius, M.T.   Grisaru, D.  Zanon,
 \NP {\bf B382} (1992) 365}

\refdef\rDDa\DDa{C.\ts Destri and H.\ts J.\ts de Vega,
 {\it Phys. Lett.} {\bf B233} (1989) 336}

\refdef\rFen\Fen{P.\ts Fendley, Preprint BUHEP-93-10}

\refdef\rFKa\FKa{A.\ts Fring and R.\ts K\"oberle, S\~ao Paulo
preprint,
USP-IFQSC/TH/93-06}

\refdef\rFKb\FKb{A.\ts Fring and R.\ts K\"oberle, S\~ao Paulo
preprint,
USP-IFQSC/TH/93-12}

\refdef\rGZ\GZ{S.\ts Ghoshal and A.\ts \Zam, Rutgers preprint RU-93-20}

\refdef\rKSS \KSS {P.\ts P.\ts Kulish, R. \ts Sasaki  and C. \ts Schwiebert,
{\it J.~Math.~Phys.} {\bf 34} (1993) 286}

\refdef\rKS\KS{P. P. \ts Kulish and R. \ts Sasaki, {\it Prog. \ts
Theor. \ts Phys.} {\bf 89} (1993) 741}

\refdef\rMajb \Majb {S.\ts Majid, {\it  J. Math. Phys.} {\bf 32} (1991)
3426, {\bf 34} (1993) 2045}

\refdef\rMNb \MNb {L.\ts Mezincescu  and R. \ts Nepomechie,
in `Quantum Groups' T.\ts Curtright, D.\ts Fairlie  and
C.\ts Zachos  (Eds.), World Scientific (1991) 206}

\refdef\rMOPa\MOPa{A. V. Mikhailov, M. A. Olshanetsky and
A. M. Perelomov,
 {\it Comm. Math. Phys.} {\bf 79} (1981) 473}

\refdef\rOTa\OTa{D. I. Olive and N. Turok,
\NP {\bf B215} (1983) 470}

\refdef\rOTc\OTc{D. I. Olive and N. Turok,
{\it Nucl. Phys.} {\bf B265}
(1986) 469}

\refdef\rOTb\OTb{D. I. Olive and N. Turok,
{\it Nucl. Phys.} {\bf B257} (1985) 277}

\refdef\rSklc \Sklc {E.\ts K.\ts Sklyanin, {\it J. Phys.}
{\bf A21} (1988) 2375}
\refdef\rSkld \Skld {P.\ts P.\ts Kulish  and E.\ts K.\ts Sklyanin, {\it J.
Phys. \bf  A25}
(1992) 5963}


\rightline{YITP/U-93-33}
\medskip
\centerline{\bf Reflection Bootstrap Equations for Toda Field Theory
\foot{Invited Plenary talk at the International Conference on
Interface between Physics and Mathematics IPM'93. Sept. 6--17, 1993
Hangzhou,China.}
}
\bigskip
\centerline{ Ryu Sasaki}
\bigskip
\centerline{Uji Research Center}
\centerline{Yukawa Institute for Theoretical Physics}
\centerline{Kyoto University, Uji 611, Japan}
\noindent{\bf Abstract}

An algebraic approach to integrable quantum field theory with a
boundary (a half line) is presented and  interesting algebraic
equations,
Reflection equations (RE) and Reflection Bootstrap equations (RBE)
are discussed.
The Reflection equations are a consistent generalisation of
Yang-Baxter equations for factorisable scatterings on a half line (or
with a reflecting boundary).
They determine the so-called reflection matrices.
However, for Toda field theory and/or
other theories with diagonal S-matrices, the Reflection-Bootstrap
equations proposed by Fring and K\"oberle
determine the reflection matrices, since the reflection
equations and the Yang-Baxter equations become trivial in these cases.
The explicit forms of the reflection matrices together with their
symmetry properties are given for various Toda field theories, simply
laced and non-simply laced.

\newsec{Introduction}

This is the second talk on Toda field theory  in this Conference.
The first part, given by Ed Corrigan, showed that Toda field theory is
a simple and very good example of $1+1$ dimensional quantum field
theory that can be solved exactly. As is well known, the quantum field
theory describes the behaviour of elementary particles, the
fundamental building blocks of the matter.
However, because of various difficulties, in particular, the infinite
degrees of freedom involved, the complete solution of quantum field
theory in $3+1$ dimensions does not seem to be an easy goal to be
achieved.
In most cases our understanding is limited to the perturbative regime.
Therefore, examples of completely solvable quantum field theories,
like the Toda field theory, in spite of its $1+1$ dimensionality, are
expected
to clarify the structure of quantum field theory beyond perturbation.
This is the general motivation of our working in Toda field theory and
its generalisation.

This second part deals with integrable quantum field theory on a half
line,
instead of the entire line. Namely, with certain boundary conditions
at the end of the half line.
This problem is also related with integrable statistical lattice
models with non-trivial boundary conditions, scattering of
electrons by an impurity in solids (the Kondo problem) and
deformations of conformal field theory with boundaries.
Here we would try to show that Toda field theory again provides a very
good theoretical laboratory.

Affine Toda field
theory \NRF\rMOPa{\MOPa\semi\OTa}\refs{\rMOPa} is a
massive scalar field theory with exponential interactions in $1+1$
dimensions
described by the Lagrangian density
\eqn\ltoda{{\cal L}={1\over 2}
\partial_\mu\phi^a\partial^\mu\phi^a-V(\phi )}
in which
\eqn\vtoda{V(\phi )={m^2\over
\beta^2}\sum_0^rn_ie^{\beta\alpha_i\cdot\phi}.}
Here $\phi$ is an $r$-component scalar field, $r$ is the rank of a
compact semi-simple Lie algebra $g$ with $\alpha_i$;
$i=1,\ldots,r$ being its simple roots.
An additional root, $\alpha_0=-\sum_1^rn_i\alpha_i$ is an integer
linear combination of the simple roots, is called the affine root;
 it corresponds to the extra spot on an extended Dynkin-Kac diagram
for $\hat g$ and $n_0=1$.
When the term containing the extra root is removed, the theory becomes
conformally invariant (conformal Toda field theory).
The simplest affine Toda field theory, based on the simplest Lie algebra
$a_1$, the algebra of $su(2)$, is called sinh-Gordon theory, a cousin
of
the well known sine-Gordon theory. $m$ is a real parameter setting
the mass scale of the theory and $\beta$ is a real coupling constant,
which is relevant only in quantum theory.

Toda field theory is integrable at the classical level due to
the presence of an infinite number of conserved quantities.
Many beautiful properties of Toda field theory, both at the classical
and quantum levels, have been explained in some detail in Ed
Corrigan's talk.
In particular, it is firmly believed that the integrability survives
quantisation.
The exact quantum S-matrices are known
\NRF\rAFZa\AFZa
\NRF\rBCDSa\BCDSa\NRF\rBCDSb\BCDSb\NRF\rBCDSc\BCDSc\NRF
\rCMa{\CMa\semi\CMb}\NRF\rDDa\DDa
\NRF\rDGZc\DGZc \NRF\rCDS\CDS
\refs{\rAFZa - \rCDS}
 for all the
Toda field theories based on non-simply laced algebras as well as those
based on simply laced algebras.
The singularity structure of the latter S-matrices, which in some
cases contain poles up to 12-th order \refs{\rBCDSc},
is  beautifully explained in terms of the
singularities of the corresponding Feynman diagrams
\NRF\rBCDSe\BCDSe\refs{\rBCDSe}, so called Landau singularities.

In this talk we need only the most rudimentary facts of Toda field theory,
namely masses and three point couplings.
Expanding the potential $V(\phi )$ around the classical vacuum
$\phi \equiv 0$
we get
\eqn\vexp{V(\phi ) = {m\sp 2\over\beta\sp 2}h + {1\over2}(M\sp 2)\sp {ab}
\phi\sp a\phi\sp b
+ {1\over3!}c\sp {abc}\phi\sp a\phi\sp b\phi\sp c + \cdots,}
where $h=\sum_0\sp r n_i$ is the Coxeter number of $g$ ,
$M\sp 2$ is the
mass matrix
$ (M\sp 2)\sp {ab} = m\sp 2\sum_i n_i\alpha_i\sp a\alpha_i\sp b .$
We choose the representation of the simple roots such that the
mass matrix becomes diagonal. Then  we get the mass spectrum as the
eigenvalues $m_1^2 , m_2^2,\ldots m_r^2$.
In this basis the cubic term in the field operators,
$c\sp  {abc}=m\sp  2\beta\sum_{i=0}\sp  r n_i
\alpha_i\sp  a\alpha_i\sp  b\alpha_i\sp  c ,$
describes the three point interaction in which particle $a$, $b$ and
$c$ meet
at one space-time point and annihilate.
The same term describes the other processes, $a$ and $b$ meet to
produce $\bar c$ (the anti-particle of $c$),
$b$ and $c$ meet to produce $\bar a$, etc.
This is due to a simple fact in relativistic field theory that a
field operator $\phi_a$ annihilates a particle $a$ and
creates an anti-particle $\bar a$. Crossing symmetry is the expression of this
fact at the S-matrix level.
Armed with these information one can go on to determine the S-matrices of
affine Toda field theory.

Let us first look at the exact factorisable S-matrices in a broader
context.
The stage is the 2 dimensional Minkowski space,
with the particles satisfying the mass-shell conditions:
$ p^2 = (p_0)^2 - (p_1)^2 = m^2,$
which can be conveniently parametrised in terms of {\it rapidity} $\theta$,
$ p_0= m\cosh\theta$, $p_1= m\sinh\theta.$
Most known examples  of exact S-matrices are obtained as solutions of the
Yang-Baxter equation, which is the necessary condition for the factorisability
of the three-body elastic S-matrix into a product of three two-body
S-matrices.
Namely the Yang-Baxter equation requires the equality of the products of
S-matrices in two different orders.
\eqn\YaBa{ S_{12}(\theta_1 -\theta_2)S_{13}(\theta_1 -\theta_3)S_{23}(\theta_2
-\theta_3) =
S_{23}(\theta_2 -\theta_3)S_{13}(\theta_1 -\theta_3)S_{12}(\theta_1
-\theta_2).}
The S-matrices depend on the rapidity difference as required by the Lorentz
invariance.

In the simplest case when the masses of the particles are all degenerate,
say $N$ fold, the S-matrices have $N^2\times N^2$ entries.
The explicit indices dependence was omitted because it would be messy and
we are not going to use it in the Toda field theory context, either.
In Toda field theory most of the particle masses are distinct, and in the less
common cases of degeneracy, these particles are distinguished by the infinite
number of conserved quantities.
Therefore, the S-matrices are diagonal, to be denoted by $S_{ab}(\theta)$
($a, b= 1, \ldots, r$ are the label of two incoming particles)
which are complex numbers with unit modulus for real $\theta$.
And the Yang-Baxter equations become trivial, since the diagonal matrices
always commute.
There is, however, another set of algebraic equations, called Bootstrap
equations which govern the exact S-matrices for Toda field theory instead of
the
Yang-Baxter equations.

Suppose $S_{ab}(\theta)$ has a pole at $\theta=i\theta_{ab}^c$
($0<\theta_{ab}^c<\pi$), corresponding to a bound state particle
$c$ with the mass
$ m_c^2 = (p_a + p_b)^2
= m_a^2 + m_b^2 + 2 m_a m_b\cos\theta_{ab}^c$ .
In Toda field theory based on simply laced algebras, this happens when and
only when the corresponding three point coupling $c\sp{ab\bar c}$ \vexp\
 is non-vanishing. (For the generalised bootstrap principle for the non-simply
laced theory, see \refs{\rCDS}.)
Then on top of the pole (ie, $\theta=i\theta_{ab}^c$) the two particle state
$ab$ is completely dominated by
the single particle state $c$, which leads to the following equation
among the S-matrices
\eqn\Boots{S_{cd}(\theta) = S_{ad}(\theta +i\bar\theta_{ac}^b)
                            S_{bd}(\theta -i\bar\theta_{bc}^a),}
in which $d$ is an arbitrary particle and
$$\theta_{ab}^c + \theta_{bc}^a + \theta_{ca}^b=2\pi, \qquad
\bar\theta = \pi -\theta,\qquad
\bar\theta_{ab}^c + \bar\theta_{bc}^a + \bar\theta_{ca}^b= \pi. \qquad
$$
Based on the mass spectrum \foot{See, for example \rBCDSc.},
 analyticity (in $\theta$),
\eqn\unit{{\rm Unitarity}\qquad S_{ab}(\theta)S_{ab}(-\theta)=1 ,}
which simply means $|S_{ab}(\theta)|=1$ for real $\theta$ and
\eqn\cros{{\rm Crossing~~ Symmetry}\qquad S_{ab}(i\pi -\theta)
                                      = S_{a\bar b}(\theta),}
together with certain conditions imposed by the Langangian
\ltoda, these over-determined set of Bootstrap equations determine
the Toda S-matrices. They are $2\pi i$ periodic function of $\theta$
and have symmetry properties due to {\bf C}, {\bf P} and {\bf T} invariance
\eqn\cpt{S_{ab}(\theta)=S_{ba}(\theta)=S_{\bar a\bar b}(\theta).}

The basic building block of the S-matrix
satisfying the unitarity  may be taken to be of the form
\eqn\block{(x)= (x)_{\theta} ={\sinh\left({\theta\over 2}+
{i\pi x\over 2h}\right)\over
\sinh\left({\theta\over 2}-{i\pi x\over 2h}\right)},}
where $h$ is the Coxeter number.
In order to incorporate the coupling constant dependence it is
advantageous to introduce the modified building block \refs\rBCDSc
\eqn\cblock{ \d{x}=\d{x}_{\theta}={(x+1)(x-1)\over (x+1-B)(x-1+B)},}
where $B=B(\beta)={{\beta^2/2\pi }\over{1+\beta^2/4\pi}}$ gives
the coupling dependence
\NRF\rBSa\BSa\refs\rBSa.
We usually suppress the $\theta$ dependence of $(x)$ and $\d{x}$
unless otherwise stated.
Here we list the S-matrices of $a_n^{(1)}$ and $d_n^{(1)}$ Toda field
theories in the above notaion:

\eqn\asfull{a_n^{(1)}:\quad
S_{ab}=\d{a+b-1}\d{a+b-3}\dots\d{\vert a-b\vert +1}
=\prod^{a+b-1}_{\vert a-b\vert +1 \atop{\rm step}\ 2}\d{p},}
\eqn\abds{d_n^{(1)}:\quad S_{ab}=\prod^{a+b-1}_{\vert a-b\vert +1
\atop {\rm step}\ 2}\d{p}\d{h-p}, \quad S_{s^\prime a}=S_{sa}=S_{\bar sa}
=\prod^{2a-2}_{ 0\atop {\rm step}\ 2}\d{n-a+p},}
\eqn\sseven{d_{\rm even}^{(1)}:\quad S_{ss}=S_{s^\prime s^\prime}
=\prod^{h-1}_{1\atop {\rm step}\ 4}\d{p}, \qquad S_{ss^\prime}
=\prod^{h-3}_{3\atop{\rm step}\ 4}\d{p},}
\eqn\ssodd{d_{\rm odd}^{(1)}:\quad S_{ss}
=\prod^{h-3}_{ 1\atop {\rm step}\ 4}\d{p}, \qquad S_{s\bar s}
=\prod^{h-1}_{ 3\atop {\rm step}\ 4}\d{p}.}

For the S-matrices of $e_n$ series and non-simply laced theories, see
\refs{\rAFZa- \rCDS}.

\newsec{Reflection equations and Reflection Bootstrap equation}

Next let us consider the factorisable scattering on a {\it half line}
$x\leq0$. This problem was first formulated by Cherednik
\NRF\rChe\Che\refs{\rChe}.
Here we have an additional element, the reflection at the boundary $x=0$,
denoted by the K-matrix, $K(\theta)$ (an $N\times N$ matrix for an
$N^2\times N^2$ S-matrix).
The K-matrices should satisfy the so-called Reflection equations
encoding the equality of the two alternative ways of reflection
\eqn\refeq{
S(\theta_1 -\theta_2)K_1(\theta_1)S(\theta_1+\theta_2)K_2(\theta_2)
=
K_2(\theta_2)S(\theta_1 +\theta_2)K_1(\theta_1)S(\theta_1-\theta_2) . }

At this point some general remarks are in order.
It is well known that the Yang-Baxter equations can be generalised
into quadratic algebras or exchange algebras, which are sometimes
called quantum groups. Their rich and beautiful
mathematical structures play important roles.
The Reflection equations can also be generalised to define quadratic
algebras (REA, Reflection equation algebras) associated with quantum
groups.
The Reflection equations and REA inherit many nice properties from
Yang-Baxter equations and quantum groups
\NRF\rSkld{See for example\semi\Skld\semi \KSS\semi \Sklc\semi
\MNb\semi \CG \semi\Majb\semi \KS\semi \DeVR\semi \Fen\semi  \AKR}
\refs\rSkld.

In the Toda field theory context or in any theory with {\it diagonal}
S-matrices, the Reflection equation is trivial.
When a particle $a$ hits the end point $x=0$ with rapidity $\theta$
it is reflected elastically ($\theta \to -\theta$) and acquires a
factor $K_a(\theta)$.
It is expected that the analog of the Bootstrap eqations,
the so-called Reflection-Bootstrap equations (RBE) determine the K-matrices.
\NRF\rFKa\FKa \NRF\rGZ\GZ
This problem was first clarified by Fring and K\"oberle \refs{\rFKa}
\eqn\refb{ K_c(\theta) = K_a(\theta+ i\bar\theta_{ac}^b)
K_b(\theta -i\bar\theta_{bc}^a) S_{ab}(2\theta +
i\bar\theta_{ac}^b-i\bar\theta_{bc}^a).}
The unitarity comditions
are  given by
\eqn\runit{{\rm Unitarity}\qquad  K_a(\theta)K_a(-\theta) = 1,}
and the  crossing conditions are derived
 by Ghoshal and Zamolodchikov\refs{\rGZ}
\eqn\rcros{{\rm Crossing~~Condition}\qquad K_a(\theta)K_{\bar
a}(\theta-i\pi) = S_{aa}(2\theta) =S_{\bar a \bar a}(2\theta).}
 It should be remarked that the crossing conditions \rcros\ need not
be imposed independently.
They are built in automatically as the consistency conditions among
the Reflection Bootstrap equation \refb\ when all the possible fusings
are taken into account, namely $a b\to c$, $b \bar c\to \bar a$ and
$\bar c  a\to \bar b$. The crossing symmetry at the field operator
level requires the latter two fusings ($b \bar c\to \bar a$ and
$\bar c a\to \bar b$) when the former $a b\to c$ exists.
However, the converse is not true.
There is no guarantee for $K$'s satisfying the Crossing conditions
\rcros\ to satisfy the Reflection Bootstrap equations \refb.

Here we list some important properties of the Reflection Bootstrap
equations.
First by comparing the Bootstrap equation and the Reflection Bootstrap
equation
\eqn\RKeq{\eqalign{S_{cd}(\theta) &= S_{ad}(\theta +i\bar\theta_{ac}^b)
                            S_{bd}(\theta -i\bar\theta_{bc}^a),\cr
 K_c(\theta) &= K_a(\theta+ i\bar\theta_{ac}^b)
K_b(\theta -i\bar\theta_{bc}^a) S_{ab}(2\theta +
i\bar\theta_{ac}^b -i\bar\theta_{bc}^a), \cr}}
it is easy to see a new solution $K\sp\prime_c(\theta)$ can be
obtained from an old one
\eqn\ssym{
K\sp\prime_c(\theta) = K_c(\theta)(S_{cd}(\theta))^{\pm1},\quad d :
\quad{\rm arbitrary}.}
In fact, any quantity satisfying the Bootstrap equation will do the
same job. (For example, the non-minimal part of the Toda S-matrix with
an arbitrary coupling dependence.)
Because of the {\bf C}, {\bf P}, {\bf T} invariance \cpt, $K\sp\prime_a(\theta)
=
K_{\bar a}(\theta)$, $a= 1,\ldots, r$ is also a solution.
Similarly
\eqn\consol{K\sp\prime_a(\theta) = K_a(\theta-i\pi) =
S_{aa}(2\theta)/K_{\bar a}(\theta),}
is also a solution. The second equality follows from the crossing
condition \rcros.

Next we write down the Bootstrap equations for $d=c$, $a$ and $b$,
\eqn\thrbot{\eqalign{S_{cc}(\theta) &= S_{ac}(\theta +i\bar\theta_{ac}^b)
                            S_{bc}(\theta -i\bar\theta_{bc}^a),\cr
S_{ac}(\theta) &= S_{aa}(\theta +i\bar\theta_{ac}^b)
                            S_{ab}(\theta -i\bar\theta_{bc}^a),\cr
S_{bc}(\theta) &= S_{ab}(\theta +i\bar\theta_{ac}^b)
                            S_{bb}(\theta -i\bar\theta_{bc}^a),\cr}}
and substitute the second and the third equations into the first to
get (with replacement $\theta\to2\theta$)
\eqn\rootsol{ S_{cc}(2\theta) = S_{aa}(2(\theta + i\bar\theta_{ac}^b))
                    S_{bb}(2(\theta - i\bar\theta_{bc}^a))
                 S_{ab}^2(2\theta + 2i\bar\theta_{ac}^b -2i\bar\theta_{bc}^a).}
It is easy to see that
\eqn\formalsol{
K_a(\theta)=\sqrt{S_{aa}(2\theta)}, \quad a=1,\ldots, r,}
formally satisfies the Reflection Bootstrap equations in all the Toda
field theories.
However, in the rest of this talk we will consider only the
meromorphic solutions.

\newsec{Solutions of Reflection Bootstrap equations for Toda theory}

Here we discuss some explicit solutions of Reflection Bootstrap
equations (RBE) for various Toda field theories.
At first some preparations are in order.
Since we need $S_{ab}(2\theta)$ etc, it is useful to introduce a new
block corresponding to the ordinary block $\d{x}$ at $2\theta$;
\eqn\twob{\d{x}_{2\theta} = [x/2]_{\theta}/[h-x/2]_{\theta}
=[x/2]/[h-x/2],}
in which a new ({\it half}) block $[x]_{\theta}= [x]$ is given in terms of
the elementary block $(x)$,
\eqn\halfb{ [x]_{\theta} =
{(x-{1\over2})(x+{1\over2})\over(x-{1\over2}+{B\over2})
(x+{1\over2}-{B\over2})}.}
The fusing angles, eg. $\theta_{ab}^c$ etc, in Toda theory are integer
multiples of $\pi/h$ (in non-simply laced case, $\pi/H$).
So we denote $x\pi/h$ simply $x_h$.
Among many possible solutions of RBE
 we mainly discuss the `minimal' solutions after \rGZ.
The `minimal' solutions have minimal set of (say, coupling
independent)
zeroes and poles in the physical strip ($0< {\rm Im}\theta<\pi$).
In other words we define the degree $D$ of the solution $\{K_a\}$,
\eqn\degdef{ D= \sum_{a=1}^r({\rm number~of~ poles~ and~ zeros~ of}~
K_a ) .}

\subsec{$a_1^{(1)}$ case}
The simplest Toda field theory ($h=2$), the sinh-Gordon theory,
 consists of one neutral particle and it has no three
point coupling. Therefore the Reflection Bootstrap equation is void.
But the Crossing condition \rcros\ makes sense.
The `minimal' solutions are
\eqn\aones{K_1(\theta)= [1/2],\quad [3/2]^{-1},
\qquad {\rm for}\quad S_{11}(2\theta)= [1/2]/[3/2].}
The above two solutions are related by \consol.

\subsec{$a_2^{(1)}$ case}
It has particles 1 and 2 which are hermitian conjugate to each other,
$2=\bar1$ and $h=3$. The RBE reads
\eqn\atwoe{\eqalign{ K_2(\theta)
&=K_1(\theta+i1_h)K_1(\theta-i1_h)S_{11}(2\theta), \cr
K_1(\theta)
&=K_2(\theta+i1_h)K_2(\theta-i1_h)S_{22}(2\theta). \cr}}
Eliminating $K_2$ in terms of $K_1$ and using the bootstrap properties
of S-matrices, we get
\eqn\atwoq{K_1(\theta+i2_h)K_1(\theta)K_1(\theta-i2_h)=1.}
The `minimal' solutions are ($D=2$)
\eqn\atwom{K_1(\theta)=1,\quad
K_2(\theta)=S_{11}(2\theta)=[1/2]/[5/2];
\quad K_1(\theta)=S_{11}(2\theta),\quad K_2(\theta)=1.}
 There are charge conjugation {\it even} solutions ($D=6$) related by \consol,
\eqn\atwoc{K_1 = K_2 = [1/2][3/2], \quad ([3/2][5/2])^{-1}.}

\subsec{$a_3^{(1)}$ case}
It has a pair of complex particles 1 and $3=\bar1$ and a neutral
particle 2 with $h=4$.
\eqn\athre{\eqalign{ K_2(\theta)
&=K_1(\theta+i1_h)K_1(\theta-i1_h)S_{11}(2\theta),\cr
K_3(\theta)
&=K_2(\theta+i1_h)K_1(\theta-i2_h)S_{12}(2\theta-i1_h),\cr
K_2(\theta)
&=K_3(\theta+i1_h)K_3(\theta-i1_h)S_{33}(2\theta),\cr
 K_1(\theta)
&=K_2(\theta-i1_h)K_3(\theta+i2_h)S_{23}(2\theta+i1_h).\cr}}
We eliminate $K_2$ and $K_3$ from the first two equations
in favour of $K_1$ then the next two equations give the same
nonlinear algebraic equation for $K_1$,
\eqn\athrt{K_1(\theta+i3_h)K_1(\theta+i1_h)K_1(\theta-i1_h)
K_1(\theta-i3_h)S_{11}(2\theta)/S_{13}(2\theta)= 1 .}
This reflects the fact that all the particles of $a_n^{(1)}$ Toda
theory can be obtained as bound
states of the `elementary' particle 1 {\it or} $\bar1$.
The \lq minimal' solutions are charge conjugation {\it even}
\eqn\athrs{K_1=K_3=[1/2], ([7/2]^{-1}),\quad K_2=[3/2]/[7/2],
([1/2]/[5/2]),\qquad D=5,}
which can be obtained by solving a much simpler  Crossing condition
\rcros\
for the `elementary' particle 1, $\bar1$ with $K_1=K_{\bar1}$,
\eqn\cevenq{K_1(\theta)K_1(\theta-i\pi)=S_{11}(2\theta) ,
}
whose solutions always satisfy the full equation \athrt.
 The full equation \athrt\ has both charge
conjugation {\it even} and {\it non-even} solutions.
The situation in $a_{\rm odd}^{(1)}$ ($a_{\rm even}^{(1)}$) theory is
about the same as in $a_3^{(1)}$ ($a_2^{(1)}$). The lack of `minimal'
charge conjugation {\it even} solutions in $a_{\rm even}^{(1)}$
($h={\rm odd}$) theory could be `understood' by the fact that the two
points corresponding to particle $a$ and $\bar a$ have different
colours when the spots of the Dynkin diagram are bi-coloured.

\subsec{$d_n^{(1)}$ theory}
In $d_n^{(1)}$ theory the `elementary' particles are $s$ and
$s\p$ ($\bar s$) corresponding to the (anti-) spinor
representations.
In contrast to the $a_n^{(1)}$ case, both $s$ {\it and} $s\p$ ($\bar s$)
are necessary to make all the particles as bound states.
Therefore in order to get nonlinear equations containing only one
unknown variable (like \atwoq\ or \athrt\ts) we need certain
assumption,
for example, $K_s=K_{s\p}$ ($K_{\bar s}$). Solutions not satisfying the
assumption must be considered separately.
Of course `non-minimal' solutions not satisfying the assumption can be
easily obtained by means of \ssym.

For concreteness, let us consider $d_4^{(1)}$, which has three mass
degenerate
neutral particles 1, $s$ and $s^\prime$ and a neutral particle 2
correponding to the central spot of the Dynkin diagram ($h=6$).
With an assumption $K_1=K_s=K_{s\p}$, the Reflection Bootstrap equation
reduce to
\eqn\dfouo{\eqalign{&K_2(\theta)=K_1(\theta+i1_h)K_1(\theta-i1_h)
S_{11}(2\theta),\cr
&K_1(\theta+i3_h)K_1(\theta-i3_h)S_{11}(2\theta)=1 ,\cr}}
\eqn\dfout{K_1(\theta)=K_1(\theta+i2_h)K_1(\theta-i2_h)S_{1s}(2\theta).}
The first eq. simply defines $K_2$ in terms of $K_1$ and the second
eq. is the crossing condition for $K_1$. The third eq. gives a
non-trivial condition.
Solutions are
\eqn\dfsol{\eqalign{
K_1&=[1/2][5/2], \quad (([7/2][11/2])^{-1}),\cr
K_2&=[3/2]^2[5/2]/[11/2], \quad ([1/2]/[7/2][9/2]^2). \cr}}
Not all the solutions of the crossing condition satisfy \dfout.

\subsec{$g_2^{(1)}$-$d_4^{(3)}$ case}
This is the simplest example of non-simply laced Toda theory
\refs{\rDGZc-\rCDS}. It has
two neutral particles, 1 and 2 with masses
\eqn\gmasses{m_1=\sin\pi /H\qquad m_2=\sin 2\pi /H ,}
up to an overall factor $(2\surd 2 m)$ which is ignored.
 The parameter $H$
floats in the range $6\le H\le 12$, ie between  the Coxeter numbers of
the partners in the pair.
It has fusing 112 and self-couplings
111 and 222. This is a good example to understand the structure of the
Reflection Bootstrap equation.
The RBE reads
\eqn\gtwoe{\eqalign{K_2(\theta)&=
K_1(\theta+i1_h)K_1(\theta-i1_h)S_{11}(2\theta),~\cr
K_1(\theta)&=K_2(\theta+i\pi-i2_H)K_1(\theta-i1_h)S_{12}(2\theta+i\pi-i3_H),\cr
K_1(\theta)&=K_1(\theta+i\pi/3)K_1(\theta-i\pi/3)S_{11}(2\theta),~\cr
K_2(\theta)&=K_2(\theta+i\pi/3)K_2(\theta-i\pi/3)S_{22}(2\theta), \cr}
}
The first eq. defines $K_2$ in terms of $K_1$, that of the
`elementary' particle.
The consistency of the first two eqs. (both derived from the same
fusing 112) gives rise to the crossing condition for $K_1$ as
mentioned before.
The crossing condition for $K_2$ is guranteed by that of $K_1$.
The third eq. implies the fourth eq. as well as the crossing condition
for $K_1$. Thus we only need solve the third eq., with solutions
\eqn\gtwos{
K_1 =[1/2]([H/2+1/2][3H/4]_{1/2})^{-1},\qquad
[H/4]_{1/2}[H/2-1/2][H-1/2]^{-1}.
}
Both are factors of
$$S_{11}(2\theta)={[1/2][H/2-1/2][H/4]_{1/2}\over
[H-1/2][H/2+1/2][3H/4]_{1/2}},
$$
 as in other cases.
Here $[H/4]_{1/2}$ and  $[3H/4]_{1/2}$
are the {\it half} blocks corresponding to a new building block
$\d{x}_{1/2}$
necessary for the non-simply laced S-matrices \refs{\rCDS},
\eqn\nublock{\{ x\}_\nu ={(x -\nu B -1)(x+\nu B +1)\over (x+ \nu
B+B-1)(x-\nu B-B+1)}.}
The bracket notation has been adjusted slightly for the
non-simply laced case and is now defined by
\eqn\Hblock{(x)={\sinh\left({\theta \over 2}+{x\pi i\over 2H}\right)\over
\sinh\left({\theta \over 2}-{x\pi i\over 2H}\right)}.}
In $g_2^{(1)}$-$d_4^{(3)}$ case $H=6 + 3B$, and $0\leq B\leq2$
but this parametrisation is not intended to
imply $B$ has the form given before for the simply laced theories.
The S-matrix is given
$$S_{11}(\theta)=\d1\d{H-1}\d{H/2}_{1/2}.$$

\newsec{Summary}

An algebraic approach to integrable quantum field theory with a
boundary (a half line) was presented with interesting algebraic equations,
Reflection equations (RE) and Reflection Bootstrap equations (RBE).
Many explicit solutions are found for the latter based on the known
exact S-matrices for Toda field theory, simply laced and non-simply laced.
There are, however, many interesting points I did not touch in
this talk.
For example: a unified description of the solutions of RBE.
Lagrangian (space-time) description of the reflection matrices and the
corresponding boundary conditions and the related problems of
translational non-invariance etc.
Explanation/interpretation of singularities and zeroes of the
reflection matrices.
Conserved quantities on a half line and relationship with deformed
conformal field theory with boundaries \refs\rGZ.
Relationship with integrable lattice lattice models with
non-trivial boundary conditions, etc.
After returning from China,  we found a second paper of Fring and
K\"oberle \NRF\rFKb\FKb \refs\rFKb\
which has some overlap with the present paper.

\bigskip
\noindent{\bf Acknowledgements}
\bigskip
We  are grateful to Ed. Corrigan, Roman Jackiw and Petr Kulish for
useful discussion.

\listrefs
\end

\epsfverbosetrue

\listrefs
\vfill\eject
\nopagenumbers
\def\epsfsize#1#2{\hsize}
\epsffile{cds1.eps}
\vfill\eject
\epsffile{cds2.eps}
\vfill\eject
\end